\documentclass[a4paper, preprint,12pt,sort&compress]{elsarticle}

\usepackage{epsfig}

\usepackage[dvipdfmx]{color}

\usepackage{times} 
\usepackage{amssymb}
\usepackage{lineno}
\usepackage{threeparttable}

\usepackage[utf8]{inputenc} 
\usepackage[T1]{fontenc} 
\usepackage{mathptmx} 

\journal{Journal of Alloys and Compounds}

\begin{document}

\begin{frontmatter}

\title{Tunable magnetic and magnetocaloric properties by thermal annealing
in ErCo$_{2}$ atomized particles}

\author[1]{Takafumi D. Yamamoto\corref{cor1}}
\ead{YAMAMOTO.Takafumi@nims.go.jp}
\author[1]{Akiko T. Saito}
\author[1]{Hiroyuki Takeya}
\author[1]{Kensei Terashima}
\author[1]{Takenori Numazawa}
\author[1,2]{Yoshihiko Takano}

\cortext[cor1]{Correpsonding author}

\address[1]{National Institute for Materials Science, Tsukuba, Ibaraki 305-0047, Japan}
\address[2]{University of Tsukuba, Tsukuba, Ibaraki 305-8577, Japan}

\begin{abstract}
Processing magnetocaloric materials into magnetic refrigerants with appropriate shapes is
essential for the development of magnetic refrigeration systems.
In this context, the impact of processing on the physical properties of magnetocaloric materials
is one of the important issues.
Here, we investigate the crystallographic, magnetic, and magnetocaloric properties of
gas-atomized particles of the intermetallic compound ErCo$_{2}$,
a giant magnetocaloric material for low-temperature applications.
The results demonstrate that
the physical properties of ErCo$_{2}$ are significantly changed by
atomization and subsequent thermal annealing.
In the as-atomized particles, the magnetic transition temperature increases from 34 to 56 K
and the phase transition changes from first order to second order.
The thermal annealing shifts the transition temperature back to the original one
and restores the first-order phase transition characteristic.
The changes in magnetic properties are closely related to
those in crystallographic properties,
suggesting the importance of the magneto-structural coupling.
The magnetic entropy change $-\Delta S_{M}$ of the particles can be tuned
in size, shape, and peak temperature depending on the annealing conditions.
The peak value of $-\Delta S_{M}$ varies in the range of 9--33 J kg$^{-1}$ K$^{-1}$
for a magnetic field change of 0--5 T.
All the ErCo$_{2}$ atomized particles have magnetocaloric properties
comparable or superior to other promising candidates
for low-temperature magnetic refrigerants.
\end{abstract}

\begin{keyword}
Magnetocaloric effect \sep
Rare-earth intermetallics \sep
Gas atomization \sep
Thermal annealing \sep
Magnetic measurements
\end{keyword}

\end{frontmatter}

\section{Introduction}
Magnetic materials with a large magnetocaloric effect (MCE)---magnetocaloric materials---
have attracted attention owing to their potential application to
highly efficient and environmentally friendly magnetic refrigeration,
which is one of the promising alternatives to conventional gas-based refrigeration techniques
\cite{Gschneidner-RPP-2005,Kitanovski-Springer-2015,Numazawa-Cryog-2014,Zhang-PhysB-2019}.
Usefulness as a magnetocaloric material is often evaluated by
the magnetic entropy change $-\Delta S_{M}$
and/or the adiabatic temperature change $\Delta T_{\rm ad}$
when an applied magnetic field varies
\cite{Tishin-JMMM-2007},
and these typically increase around the magnetic transition temperature $T_{\rm mag}$ of
the respective material.

Based on the nature of the magnetic phase transition,
magnetocaloric materials are mainly divided into two types
\cite{Smith-AEM-2012,Lyubina-JPD-2017,Gottschall-AEM-2019},
though some of which cannot be simply classified.
One is materials that undergo first-order magnetic phase transitions.
Their advantage is to have a huge MCE with latent heat,
whereas the operating temperature range is limited to the immediate vicinity of $T_{\rm mag}$.
The other is materials exhibiting second-order magnetic phase transitions.
Although the MCE size of these materials is inferior to the former materials,
a broad temperature dependence of their MCEs extends the operating temperature range,
which is beneficial for use in magnetic refrigeration systems.

The C15 Laves phase ErCo$_{2}$ compound is one of the promising magnetocaloric materials
for low-temperature applications.
The giant MCE of this compound,
$-\Delta S_{M} =$ 36.8 J kg$^{-1}$ K$^{-1}$ for a field change of 0--5 T
\cite{Wada-JAC-2001},
is the consequence of a first-order magnetostructural phase transition at 33 K
from the paramagnetic cubic phase to the ferrimagnetic rhombohedral phase upon cooling
\cite{Oesterreicher-JAP-1970,Duc-JPFMP-1988,Gratz-JMMM-1994}.
It is widely believed that this transition is closely related to
the onset of the itinerant electron metamagnetism of the Co sublattice,
induced by the exchange field acting on Co sites via the Er-Co exchange interaction
\cite{Duc-Handbook-1999,Kozlenko-SciRep-2015}.
In the ferrimagnetic state, the Co magnetic moment of no more than 1.0 $\mu_{\rm B}$/Co
couples with the larger Er magnetic moment antiferromagnetically
\cite{Moon-JAP-1965}.

It is known that the physical properties of ErCo$_{2}$ are sensitive to
various stimuli such as element substitution and pressure.
A small amount of Co-site substitution shifts
the ferrimagnetic transition temperature $T_{\rm C}$
and changes the phase transition from first order to second order
\cite{Duc-PhysC-1988,Soares-JMMM-1999,Danis-JAC-2002,Prokleska-JAC-2004,Liu-JAP-2008,Tang-NatCom-2022}.
In Er(Co$_{1-x}$Ni$_{x}$)$_{2}$, for instance,
a 5--10\% Ni substitution causes the loss of the magnetic and magnetocaloric properties
typical of first-order phase transitions
\cite{Wada-JAC-2001,Soares-JMMM-1999}.
A similar change in the magnetic phase transition is also observed
with the application of hydrostatic pressure
\cite{Kagayama-JMSJ-1999,Hauser-PRB-2000,Syshchenko-PRB-2001}.
When the pressure increases up to 2 GPa,
the $T_{\rm C}$ immediately decreases
with a concomitant decrease in the Co magnetic moment
\cite{Kozlenko-SciRep-2015,Ishimatsu-PRB-2007}.
These phenomena have been discussed in relation to the instability of the Co metamagnetism,
but the microscopic mechanism has not been clarified yet.

Meanwhile, it is desirable to process the candidate materials for magnetic refrigerants
into a spherical shape for the better performance of practical refrigeration systems
\cite{Yu-IJR-2010,Tusek-IJR-2013}.
We have been working on the fabrication of spherical magnetocaloric particles
by using an electrode induction melting gas atomization (EIGA) process
\cite{TDY-HoB2-2021,TDY-HoAl2-2022,TDY-HoB2-x-2022}.
In this process, an electrode rod of a target material is inductively melted
in a contactless manner in an inert atmosphere,
after which the freely-fallen molten metal is atomized with an inert gas.
Eventually, the target material rapidly solidifies into spheres
during flight in the sample chamber.
When applying the EIGA process to ErCo$_{2}$,
it is not self-evident how such a quenching effect affects
the stimulus-sensitive physical properties,
being of interest from both scientific and engineering perspectives.
In this paper, we investigate ErCo$_{2}$ gas-atomized particles
in terms of crystal structure, magnetism, and magnetocaloric effect.
The annealing effect is examined as well.

%
%
\section{Material and methods}
ErCo$_{2}$ ingots were made as the starting material
by high-frequency melting a stoichiometric mixture of Er and Co chunks.
Electrode rods used in the EIGA process were prepared by
sintering the crashed ingots in a vacuum at 1273--1473 K for 20--24 h.
After the atomization process,
the obtained atomized powder was sieved and divided into
spherical particles and irregular ones.
The former were used for measurements.
Details of the preparation of atomized particles are described elsewhere
\cite{TDY-HoB2-2021,TDY-HoAl2-2022}.
To prepare an annealed ingot as a reference sample,
a portion of the as-cast ingot was wrapped with a Ta sheet,
sealed in a SUS tube in an Ar atmosphere,
then annealed at 1073 K for 120 h.

Powder X-ray diffraction (XRD) measurements with Cu K$\alpha$ radiation were performed
at room temperature by a Rigaku MiniFlex600 diffractometer.
The temperature ($T$) and magnetic field ($\mu_{0}H$) dependence of
the magnetization ($M$) were measured by a Quantum Design SQUID magnetometer.
In the measurements on the atomized particles,
20--25 particles with a diameter of 212--355 $\mu$m were
in contact with each other and arranged vertically long.
A magnetic field was applied along the longitudinal direction of the arranged particles.
Demagnetization field correction was not performed for all the data presented here
because it is difficult to determine the demagnetization factor
for a group of the particles.
The temperature dependence of the specific heat ($C$) at 0 T was measured
by a thermal relaxation method using a Quantum Design PPMS.
In order to properly obtain the specific heat data near
the first-order ferrimagnetic magnetic transition of ErCo$_{2}$,
we referred to the paper that introduces the analysis method of
the measurement data taken by relaxation calorimetry
for materials with first-order phase transitions
\cite{Suzuki-Cryog-2010}.
%
\section{Results and discussion}
\subsection{Impacts of the atomization process}
\begin{figure}[t]
\centering
\includegraphics[width=120mm]{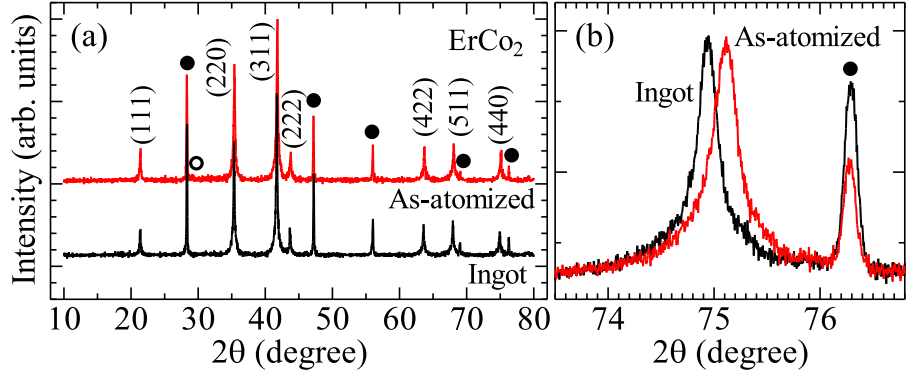}
\caption{(Color Online)
(a) Powder XRD patterns of the as-atomized particles and the annealed ingot.
The closed and open circles represent the Bragg peaks of
the internal standard Si and the impurity Er$_{2}$O$_{3}$.
(b) XRD patterns of the two samples near the (440) cubic Bragg peak.}
\label{fig:XRD}
\end{figure}

Figure \ref{fig:XRD}(a) shows powder XRD patterns of
the as-atomized particles and the annealed ingot at room temperature.
For these measurements, Si powder was mixed into each sample
as an internal standard sample,
whose the Bragg peaks are indicated by the closed circles.
Both the two patterns can be indexed by the Laves phase cubic structure
with a space group of \textit{Fd$\bar{3}$m}.
As can be seen from Fig. \ref{fig:XRD}(b), 
the (440) cubic Bragg peak of the as-atomized particles shifts
in 2$\theta$ upward 0.17$^{\circ}$ compared to the annealed ingot,
suggesting the lattice contraction.
Moreover, the Bragg peak is broader in the as-atomized particles;
the full width at half maximum (FWHM) of the (440) peak are
0.30$^{\circ}$ for the as-atomized particles
and 0.24$^{\circ}$ for the ingot sample, respectively.
These results imply that the atomization process alters
the crystallographic parameters of ErCo$_{2}$,
probably due to the rapid quenching.
Besides, the small peak near 29$^{\circ}$ in the as-atomized particles,
indicated by the open circle in Fig. \ref{fig:XRD}(a),
is due to Er$_{2}$O$_{3}$.
This oxide with a tiny amount should not affect
the physical properties of the ferrimagnetic ErCo$_{2}$
as it is an antiferromagnet with a N$\acute{\rm e}$el temperature of 3.3 K
\cite{Tang-PRB-1992}.

\begin{figure}[t]
\centering
\includegraphics[width=110mm]{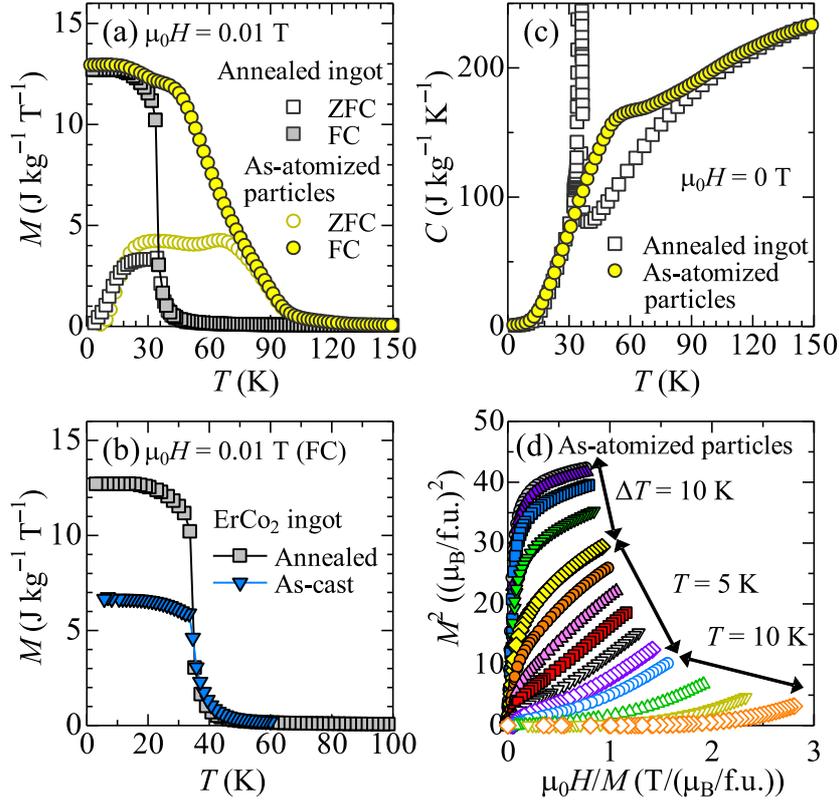}
\caption{(Color Online)
(a) $M$-$T$ curves of the ErCo$_{2}$ annealed ingot
and the ErCo$_{2}$ as-atomized particles at 0.01 T in ZFC and FC processes.
(b) The comparison of $M$-$T$ curves at 0.01 T in FC processes
between the as-cast ingot and the annealed ingot.
(c) Specific heat of the annealed ingot and the as-atomized particles at 0 T.
(d) Arrott plots at various temperatures for the as-atomized particles.
$\Delta T$ indicates the temperature interval of the measurements.}
\label{fig:Asatomized}
\end{figure}

Figure \ref{fig:Asatomized}(a) shows the isofield magnetization ($M$-$T$) curves of 
the as-atomized particles and the annealed ingot
measured at 0.01 T between 2 K and 150 K
in zero-field (ZFC) and field cooling (FC) processes.
The ingot sample exhibits a sudden increase in magnetization in the FC process,
being ascribed to the first-order ferrimagnetic transition.
The transition temperature $T_{\rm C}$,
defined as a peak temperature of the temperature derivative $|$d$M$/d$T$$|$,
is evaluated to be 34 K,
which nearly coincides with the literature data
\cite{Wada-JAC-2001,Duc-JPFMP-1988,Danis-JAC-2002}.
At temperatures below $T_{\rm C}$, a thermal hysteresis is observed
between the ZFC and FC curves.
$M$-$T$ curves are quite different for the as-atomized particles:
the ZFC curve exhibits two smooth peaks at 33 K and 65 K;
$M$ in the FC process increases slowly over a wide temperature range from 100 to 40 K,
followed by an additional step-like increase below 33 K;
and a thermal hysteresis spreads between 2 K and 90 K.
Furthermore, $|$d$M$/d$T$$|$ takes a maximum value at about 56 K.
This temperature could be regarded as the $T_{\rm C}$ of the as-atomized particles.
This point will be addressed in the following section.

The as-cast ingot sample exhibits a similar $M$-$T$ curve to the annealed one,
with minor differences,
as shown in Fig. \ref{fig:Asatomized}(b).
This result indicates that the atomization process does have a significant role in
modyfying the magnetic properties of ErCo$_{2}$.
Dramatic changes after atomizing can also be found in
the specific heat shown in Fig. \ref{fig:Asatomized}(c).
For the ingot sample, a spike-like anomaly is observed near 34 K at 0 T,
which is a typical feature of the first-order phase transition.
In stark contrast, the as-atomized particles exhibit
a broad specific heat anomaly between 40 K and 70 K.
As shown in Fig. \ref{fig:Asatomized}(d),
the Arrott plots show positive slopes at any temperatures from 10 K to 110 K,
suggesting that the magnetic transition is of second order
\cite{Banerjee-PhysLett-1964}.
Therefore, it is found that the EIGA process changes
the magnetic transition of ErCo$_{2}$ from first order to second order.

\subsection{Annealing effect on the magnetic properties}
Figures \ref{fig:XRD} and \ref{fig:Asatomized} imply that
the rapid quenching during the atomization process affects
the physical properties of ErCo$_{2}$
via the modification of the magneto-structural coupling.
In general, the effects of quenching may be removed  by thermal annealing.
Hence we performed heat treatments for the as-atomized particles
under various annealing temperatures ($T_{\rm an}$) of 473--1073 K
and annealing times ($t_{\rm an}$) of 0.5--168 h,
where the particles wrapped with a Ta sheet were sealed in a SUS tube,
annealed in an Ar atmosphere,
and then quenched into room-temperature water.
Powder XRD measurements reveal that after annealing,
the Bragg peak of Er$_{2}$O$_{3}$ does not increase
and no additional impurity phase is detected (Fig. S1).

\begin{figure}[t]
\centering
\includegraphics[width=130mm]{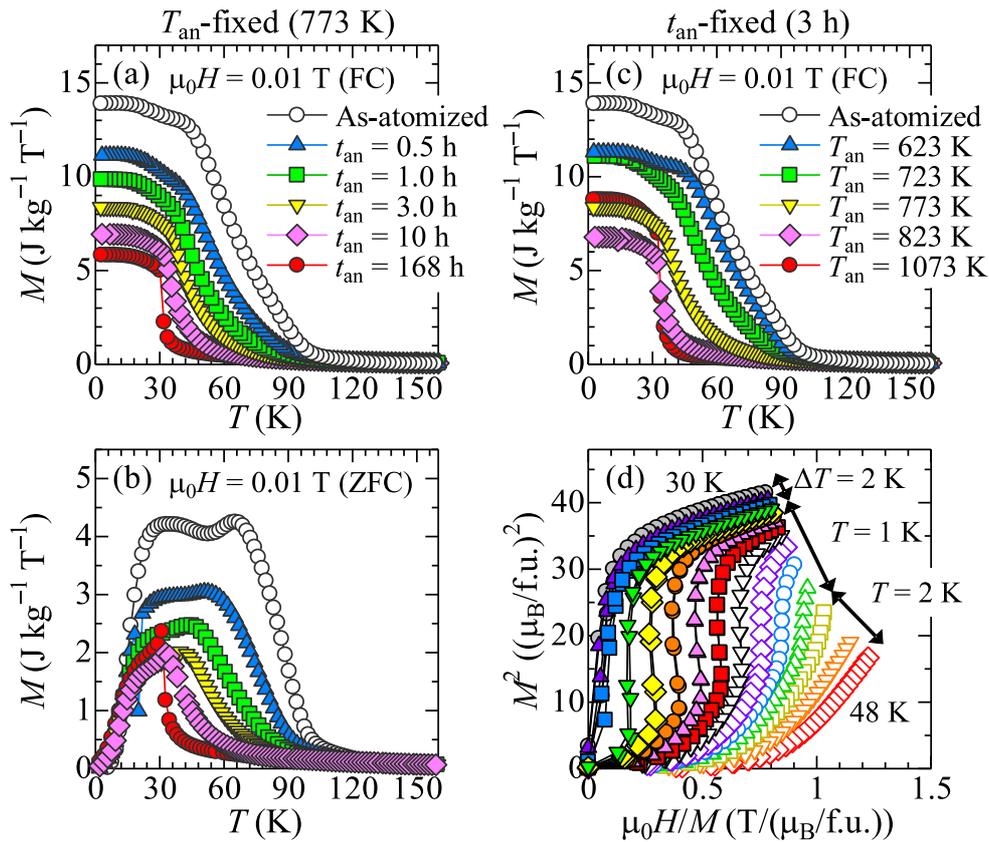}
\caption{(Color Online) (a)-(b) $M$-$T$ curves at 0.01 T in FC (a) and ZFC (b) processes
for the atomized particles in the $T_{\rm an}$-fixed case (see the text).
The legend is common to (a) and (b).
(c) $M$-$T$ curves at 0.01 T in FC processes for the atomized particles
in the $t_{\rm an}$-fixed case (see the text).
(d)Arrott plots at various temperatures near the magnetic transition tempearture
for the atomized particles annealed at 1073 K for 3h.
$\Delta T$ indicates the temperature interval of the measurements.
}
\label{fig:Magnetic}
\end{figure}

Figures \ref{fig:Magnetic}(a) and \ref{fig:Magnetic}(b) show
$M$-$T$ curves measured at 0.01 T in FC and ZFC processes
for the atomized particles annealed at $T_{\rm an} =$ 773 K
for several $t_{\rm an}$ ($T_{\rm an}$-fixed case).
In the FC processes, the longer the annealing time,
the lower the temperature at which the magnetization rises
and the sharper the rise.
Correspondingly, the peak at 65 K observed in the ZFC processes shifts to 31 K with sharpening.
The peak at 33 K also shifts to a lower temperature
and is eventually observed as a shoulder structure near 18 K.
As shown in Fig. \ref{fig:Magnetic}(c),
the same effect can be achieved by
annealing at various temperatures for $t_{\rm an} =$ 3 h ($t_{\rm an}$-fixed case).
The thermal annealing has little effect on the magnetic properties
when $T_{\rm an}$ is 623 K or less,
whereas the $M$-$T$ curve changes drastically with further increasing $T_{\rm an}$.
The change seems to mostly finish at $T_{\rm an} =$ 823 K,
above which the positions of the $M$-$T$ curve are nearly the same for any $T_{\rm an}$.

Comparing Figs. \ref{fig:Magnetic}(a)-\ref{fig:Magnetic}(c) to Fig. \ref{fig:Asatomized},
one can find that the $M$-$T$ curves of the well-annealed samples resemble
those of the ErCo$_{2}$ ingot sample.
This fact suggests that the thermal annealing restores
the first-order magnetic transition of ErCo$_{2}$.
Figure \ref{fig:Magnetic}(d) shows the Arrott plots of
the atomized particles annealed at 1073 K for 3 h.
Negative slopes are clearly observed near the magnetic transition temperature,
also supporting the change in magnetic phase transition from second order to first order
while annealing in the atomized particles
\cite{Banerjee-PhysLett-1964}.

\begin{figure}[t]
\centering
\includegraphics[width=80mm]{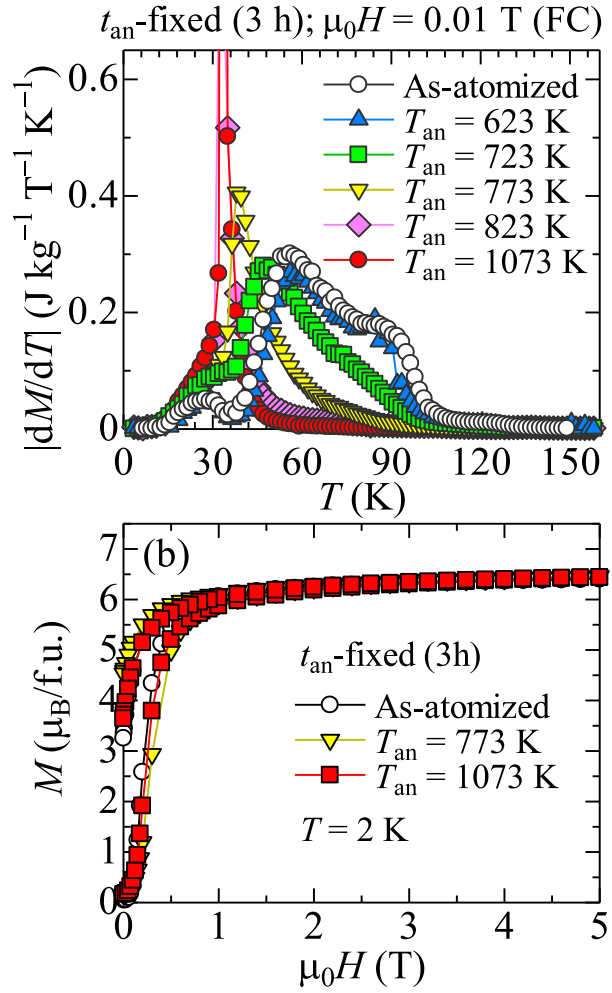}
\caption{(Color Online) (a) Temperature dependence of
the temperature derivative of magnetization |d$M$/d$T$|
at 0.01 T in FC processes for the atomized particles
in the $t_{\rm an}$-fixed case (see the text).
(b) $M$-$H$ curves at 2 K for the several particles
in the $t_{\rm an}$-fixed case (see the text).}
\label{fig:Transitions}
\end{figure}

Figure \ref{fig:Transitions}(a) shows the temperature dependence of $|$d$M$/d$T$$|$
calculated from Fig. \ref{fig:Magnetic}(c).
For the as-atomized particles, three peaks are observed:
the smallest one around 27 K, the largest one around 56 K,
and the medium-sized one around 87 K.
Among them, the largest peak systematically changes
its shape and position by thermal annealing;
its peak temperature naturally reaches the original $T_{\rm C}$ of ErCo$_{2}$.
Therefore, it seems that the peak temperature of the maximum peak can be defined
as $T_{\rm C}$ of the atomized particles.
However, this peak is broad for samples that are not well-annealed,
implying that the magnetic transitions in these samples are somewhat smeared
and might have a distribution of $T_{\rm C}$.
In that sense, it would be better not to consider the obtained $T_{\rm C}$ for these samples
as a well-defined magnetic transition temperature.
Meanwhile, as $T_{\rm an}$ increases,
the medium-sized peak shifts to lower temperatures and eventually disappears.
The smallest peak becomes smeared while keeping peak temperature
and less visible corresponding to
the disappearence of the small step-like increase of the $M$-$T$ curve.

The nature of magnetic transitions in atomized particles remains unclear at this stage.
Nevertheless, the isothermal magnetization ($M$-$H$ curve) gives more insight.
As shown in Fig. \ref{fig:Transitions}(b),
$M$-$H$ curves at 2 K are the same for the atomized particles in the $t_{\rm an}$-fixed case,
implying that the magnetic ground state is identical for all the samples,
despite the large differences in magnetic transitions.
The saturation moment is evaluated to be 6.5 $\mu_{\rm B}$ per formula unit,
which roughly coincides with the value corresponding to the ferrimagnetic order
consisting of one Er moment of 9.0 $\mu_{\rm B}$ and two Co moments of 1.0 $\mu_{\rm B}$
\cite{Moon-JAP-1965}.
Accordingly, the magnetic transitions of atomized particles are modulated from
the original magnetic transition of ErCo$_{2}$,
but it seems that the magnetic moments of both Er and Co are still involved
in all these transitions.

\subsection{Correlation between crystallographic and magnetic properties}
\begin{figure}[t]
\centering
\includegraphics[width=120mm]{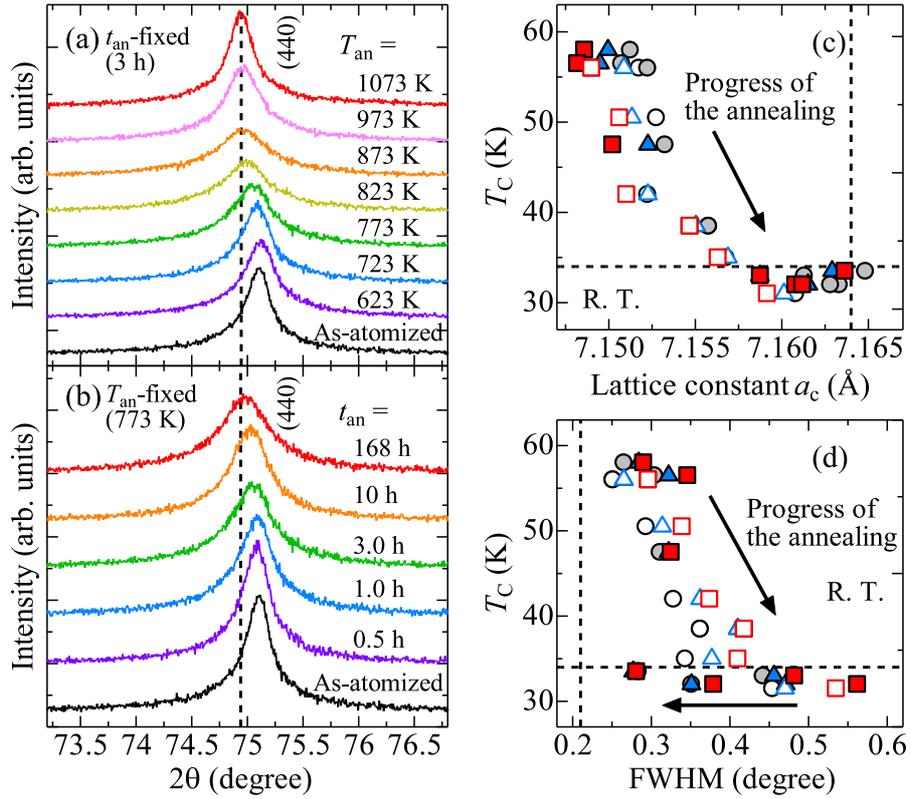}
\caption{(Color Online)
(a)-(b) Powder XRD patterns near the (440) cubic Bragg peak
in the $t_{\rm an}$-fixed case (a) and the $T_{\rm an}$-fixed case (b) (see the text).
The dotted vertical lines correspond to the peak position for the ingot sample.
(c)-(d) $T_{\rm C}$ as a function of the cubic lattice constant $a_{\rm c}$ (c)
and the full width at half maximum (FWHM) (d)
estimated from the (422) peak (circles),
the (511) peak (triangles),
and the (440) peak (squares)
in the $t_{\rm an}$-fixed case (filled symbols)
and the $T_{\rm an}$-fixed case (open symbols).
The dotted lines represent the values for the ingot sample.
}
\label{fig:Tc&Lattice}
\end{figure}

Figure \ref{fig:Tc&Lattice}(a) shows the powder XRD patterns of ErCo$_{2}$ atomized particles
near the (440) cubic Bragg peak in the $t_{\rm an}$-fixed case.
The (440) peak begins to shift to the lower angle side
when $T_{\rm an}$ exceeds 623 K and nearly stops moving above $T_{\rm an} =$ 873 K.
This behavior is reminiscent of the shift of the $M$-$T$ curve.
The peak position of the (440) peak for $T_{\rm an} =$ 1073 K is
in good agreement with that for the ingot sample,
represented by the dotted line.
In constrast to the monotonous change in peak position,
the peak width has a non-trivial $T_{\rm an}$-dependence:
the (440) peak first becomes broader as $T_{\rm an}$ increases toward 873 K
and then sharpens again with further increasing $T_{\rm an}$.
As shown in Fig. \ref{fig:Tc&Lattice}(b),
a similar peak shift by thermal annealing also occurs in the $T_{\rm an}$-fixed case,
but the peak width here continues to increase as $t_{\rm an}$ increases.

Now we will investigate the relationship of the magnetic transition temperature $T_{\rm C}$
with the cubic lattice constant $a_{\rm c}$ and FWHM of each cubic Bragg peak.
Here $a_{\rm c}$ was estimated from three peaks with Miller indices ($hkl$) of
(422), (511), and (440) in the high-angle region
by using the following expression,
\begin{equation}
a_{\rm c} = d_{(hkl)} \sqrt{h^2 + k^2 + l^2}
= \frac{\lambda_{\rm Cu}}{2 {\rm sin} \theta_{\rm max}} \sqrt{h^2 + k^2 + l^2},
\end{equation}
where $d_{(hkl)}$ is the interplaner spacing for the ($hkl$) reflection,
$\lambda_{\rm Cu}$ is the wave length of Cu K$\alpha$ radiation,
$\theta_{\rm max}$ is the angle at which each peak reaches a maximum.
Figure \ref{fig:Tc&Lattice}(c) indicates that
$T_{\rm C}$ has a negative correlation with $a_{\rm c}$
and the value of both the two approaches the values for the ingot sample
represented by the dotted lines
as the annealing progresses.
This result evidences that the magneto-structural coupling plays an important role
in determining the magnetic properties of ErCo$_{2}$ atomized particles.
On the other hand, no one-to-one correspondence is found
between $T_{\rm C}$ and FWHM (Fig. \ref{fig:Tc&Lattice}(d)).
Nonetheless, we notice that the atomized particles have
a lower $T_{\rm C}$ than the ingot sample
when FWHM is maximized,
implying that FWHM-related crystallographic parameters,
such as crystallite size or non-uniform strain,
have a secondary effect on the magnetism.

Previous pressure experiments have revealed that
$T_{\rm C}$ of ErCo$_{2}$ decreases with the lattice contraction,
being attributed to the collapse of the Co metamagnetism
due to the modification of the electronic band structure
\cite{Syshchenko-PRB-2001,Kozlenko-SciRep-2015}.
The relationship between $T_{\rm C}$ and $a_{\rm c}$ in Fig. \ref{fig:Tc&Lattice}(c)
clearly contradicts these results,
suggesting that different mechanisms cause
the changes in physical properties by the atomization process.
Although the excact mechanism remains unclear at the present stage,
let us remember here that the material is rapidly solidified from the liquid state
during the atomizatoin process.
It is then speculated that such a quenching effect realizes
some metastable state in the as-atomized particles at room temperature.
The thermal annealing could relax the system to a thermal equilibrium state,
thereby restoring the first-order phase transition characteristic of ErCo$_{2}$.

In this context, it is worth noting that
the Er-deficient Er$_{1-x}$Co$_{2}$ have similar features to the as-atomized particles
\cite{Zou-JAC-2015}:
the increase of $T_{\rm C}$ with the lattice contraction
and the weakening of the first-order phase transition characteristic.
Nevertheless, it seems unlikely that the as-atomized particles are
in a non-stoichiometric composition
as the original physical properties are restored by thermal annealing.
One possibility is that the rapid quenching introduces frozen Er-vacancies into the sample,
but the underlying mechanism of the quenching and annealing effects on ErCo$_{2}$ needs
further investigation including more comprehensive annealing experiments
and microscopic measurements such as TEM observation.

\subsection{Magnetocaloric properties of ErCo$_{2}$ atomized particles}
In this study, we will evaluate the MCE of the atomized particles by $-\Delta S_{M}$
calculated from a bunch of $M$-$T$ curves measured
at various magnetic fields under field cooling processes
through one of the Maxwell's relations:
\begin{equation}
-\Delta S_{M}(T, \mu_{0}\Delta H)
= -\{S (T, \mu_{0}H) - S(T, 0)\}
= -\mu_{0} \int_{0}^{H} \left(\frac{\partial M}{\partial T}\right)_{H} dH.
\end{equation}
It is known that the use of this equation without special care causes
spurious results with serious overestimation of $-\Delta S_{M}$
for some materials with first-order magnetic phase transitions
\cite{Liu-APL-2007,Caron-JMMM-2009,Cui-APL-2010,Franco-JPD-2017}.
Such materials are characterized by
a stepwise behavior of the $M$-$H$ curve
and a large thermal hysteresis of the $M$-$T$ curve
near $T_{\rm mag}$.
In contrast, these features are less pronounced
even for the well-annealed ErCo$_{2}$ particles (Fig. S2),
suggesting that there is no serious problem
in evaluating $-\Delta S_{M}$ of our samples
by using Eq. (2) without a particular procedure.
In fact, Wada et al. have reported that
not much difference in $-\Delta S_{M}$ is seen for ErCo$_{2}$
whether it is calculated from the magnetization data using Eq. (2) as usual
or from the heat capacity data
\cite{Wada-JAC-2001}.

\begin{figure}[t]
\centering
\includegraphics[width=65mm]{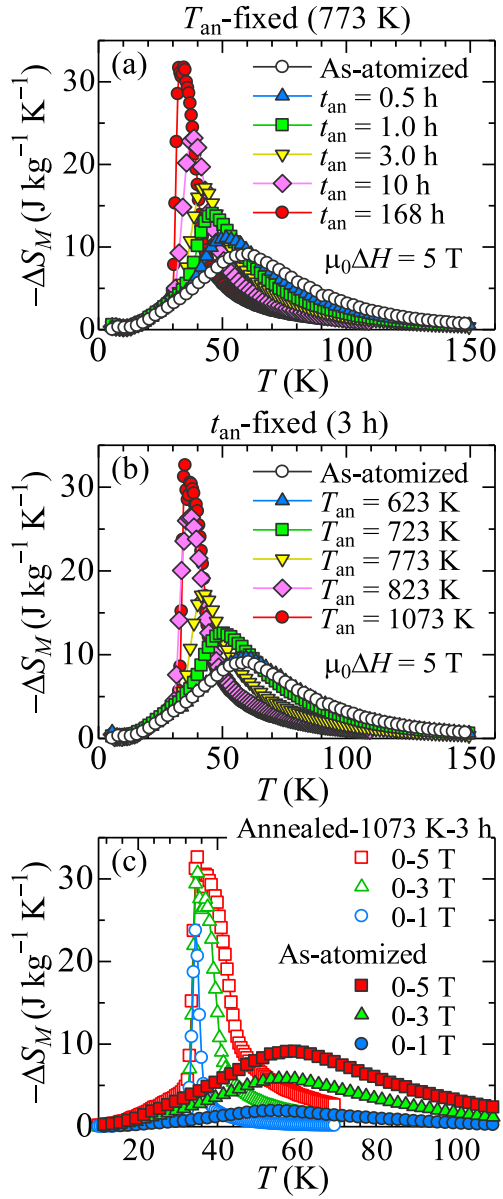}
\caption{(Color Online)
(a)-(b) Temperature dependence of $-\Delta S_{M}$ for $\mu_{0}\Delta H =$ 5 T of
the atomized particles in the $T_{\rm an}$-fixed case (a)
and the $t_{\rm an}$-fixed case (b) (see the text).
(c) $-\Delta S_{M}$ for $\mu_{0}\Delta H =$ 1, 3, and 5 T of
the as-atomized particles and the particles annealed at 1073 K for 3h.}
\label{fig:DeltaS}
\end{figure}

Figures \ref{fig:DeltaS}(a) and \ref{fig:DeltaS}(b) show
the temperature dependence of $-\Delta S_{M}$ of the ErCo$_{2}$ atomized particles
for a field change $\mu_{0}\Delta H$ of 5 T
in the $T_{\rm an}$-fixed case and the $t_{\rm an}$-fixed case, respectively.
The most striking finding is that the $-\Delta S_{M}$-$T$ curve is tunable
in terms of size, shape, and peak temperature,
depending on the annealing conditions.
In the well-annealed samples such as
the particles for $t_{\rm an} =$ 168 h and those for $T_{\rm an} =$ 1073 K,
a sharp peak of $-\Delta S_{M}$ is observed near their $T_{\rm C}$.
The peak value ($-\Delta S^{peak}_{M}$) reaches about 32--33 J kg$^{-1}$ K$^{-1}$,
comparable to the value for ErCo$_{2}$ ingot
\cite{Wada-JAC-2001}.
As the peak temperature increases,
the size of $-\Delta S_{M}$ decreases,
but the peak gradually widens.
This peak widening contributes to extend the temperature range
where $-\Delta S_{M}$ maintains large values,
which is beneficial for magnetic refrigeration applications.

As shown in Fig. \ref{fig:DeltaS}(c),
the $-\Delta S_{M}$-$T$ curve of the as-atomized particles increases in size
with increasing $\mu_{0}\Delta H$ while preserving a symmetric caret-type shape,
which is indicative of a second-order magnetic phase transition
\cite{Lyubina-JPD-2017}.
This is consistent with the conclusion drawn from Fig. \ref{fig:Asatomized}.
On the other hand, for the particles annealed at 1073 K for 3 h,
the $-\Delta S_{M}$-$T$ curve exhibits sharp, asymmetric peaks
that expand diagonally toward the higher temperature side with increasing $\mu_{0}\Delta H$.
These characteristics are typical of materials with first-order magnetic phase transitions
\cite{Lyubina-JPD-2017},
confirming that the magnetic transition is of first order
in the well-annealed particles.
Accordingly, it is found that the changes in $-\Delta S_{M}$ by thermal annealing reflect
the change in the nature of the magnetic phase transition.
For the moderately annealed particles,
the $-\Delta S_{M}$-$T$ curve with the intermediate characters is observed:
its peak expands both vertically and diagonally
as $\mu_{0}\Delta H$ increases (Figs. S3 and S4).

Table \ref{table:MCEs} summarizes the comparison of the magnetocaloric effect
between the studied materials and other promising materials
with similar magnetic transition temperatures.
The value of $-\Delta S^{peak}_{M}$ is also shown in volumetric units
because it is more meaningful from an engineering point of view
\cite{Gschneidner-RPP-2005}.
It can be seen that all the studied materials including the as-atomized particles have
the magnetocaloric properties comparable or superior to other promising materials,
suggesting their good potential as a magnetic refrigerant.
Furthermore, we would like to emphasize here that
they have different optimal operating temperature ranges.
Typically, multiple magnetocaloric materials are needed
to cover the operating temperature range required for magnetic refrigeration systems.
Taking advantage of the unique effects of atomization and thermal annealing on ErCo$_{2}$,
it is possible to prepare multiple magnetic refrigerants
that cover a relatively wide temperature range
by processing a single material.

%
\section{Conclusions}
The crystallographic, magnetic and magnetocaloric properties of
ErCo$_{2}$ gas-atomized particles have been investigated.
The atomization process changes the nature of the magnetic phase transition
from first order to second order,
and consequently the physical properties of ErCo$_{2}$ are largely modified.
The atomized particles regain
the first-order phase transition characteristic by thermal annealing,
and its magnetic transition temperature $T_{\rm C}$ returns from 56 K to about 33 K.
The change in $T_{\rm C}$ has a clear one-to-one correspondance
with the change in cubic lattice constant,
evidencing the importance of the magneto-structural coupling
on the magnetic properties of the ErCo$_{2}$ atomized particles.
Such annealing effects provide the magnetic entropy change
with tunable size (9.13--32.63 J kg$^{-1}$ K$^{-1}$, 0--5T) and operating temperature range.
The comparison with other promising materials suggests that
all the ErCo$_{2}$ atomized particles have good potential as magnetic refrigerants
for low-temperature magnetic refrigeration.
The intriguing effects of material processing presented in this study could be widely found
in other magnetocaloric materials with magnetostructural transitions.

\section*{Acknowledgements}
This work was supported by JST-Mirai Program Grant Number JPMJMI18A3, Japan.


\newpage
\begin{threeparttable}[t]
\caption{The comparison of the magnetocaloric effect for a field change of 5 T
between the ErCo$_{2}$ atomized particles
and other magnetocaloric materials with similar magnetic transition temperatures.}
\footnotesize
\begin{tabular}{llllll}
\hline
Materials & T$_{\rm mag}$ (K) & $-\Delta S^{peak}_{M}$ (J kg$^{-1}$ K$^{-1}$)
& $-\Delta S^{peak}_{M}$ (J cm$^{-3}$ K$^{-1}$)
& $\rho$ (g cm$^{-3}$) \tnote{a} & Ref.\\
\hline
ErCo$_{2}$-asatomized & 56.0 & 9.13 & 0.094 & 10.35 & Present\\
ErCo$_{2}$-623 K-3.0 h & 56.5 & 9.62 & 0.100 & 10.35 & Present\\
ErCo$_{2}$-773 K-0.5 h & 50.5 & 11.16 & 0.116 & 10.35 & Present\\
ErCo$_{2}$-723 K-3.0 h & 47.5 & 12.55 & 0.130 & 10.35 & Present\\
ErCo$_{2}$-773 K-1.0 h & 42.0 & 14.02 & 0.145 & 10.35 & Present\\
ErCo$_{2}$-773 K-3.0 h & 38.5 & 17.36 & 0.180 & 10.35 & Present\\
ErCo$_{2}$-773 K-10 h & 35.0 & 23.20 & 0.240 & 10.35 & Present\\
ErCo$_{2}$-823 K-3.0 h & 33.0 & 26.58 & 0.275 & 10.35 & Present\\
ErCo$_{2}$-773 K-168 h & 31.0 & 31.70 & 0.328 & 10.35 & Present\\
ErCo$_{2}$-1073 K-3.0 h & 33.5 & 32.63 & 0.338 & 10.35 & Present\\
\hline
ErFeAl & 55.0 & 6.1 & 0.047 & 7.72 & \cite{Zhang-Intermet-2015}\\
GdCo$_{3}$B$_{2}$ & 54.0 & 9.4 & 0.083 & 8.83 & \cite{Li-JAC-2011}\\
EuAuZn & 51.0 & 9.1 & 0.091 & 10.02 & \cite{Li-IEEETM-2014}\\
Dy$_{0.5}$Ho$_{0.5}$Al$_{2}$ & 46.0 & 21.8 & 0.132 & 6.04 & \cite{Hashimoto-ACEM-1986}\\
TbN & 44.0 & 20.5 & 0.196 & 9.57 & \cite{Yamamoto-JAC-2004}\\
Tm$_{2}$Cu$_{2}$In & 39.4 & 14.4 & 0.137 & 9.53 & \cite{Zhang-JPD-2016}\\
Ho$_{2}$Co$_{2}$Ga & 38.5 & 11.7 & 0.111 & 9.50 & \cite{Zhang-JAP-2018}\\
NdMn$_{2}$Ge$_{0.4}$Si$_{1.6}$ & 36.0 & 18.4 & 0.130 & 7.04 & \cite{Wang-APL-2011}\\
Ho$_{2}$Cu$_{2}$Cd & 30.0 & 20.3 & 0.184 & 9.08 & \cite{Yi-Intermet-2017}\\
HoAl$_{2}$ & 26.8 & 28.8 & 0.176 & 6.11 & \cite{Hashimoto-ACEM-1986}\\
\hline
\end{tabular}
\begin{tablenotes}
\item[a] The ideal density are taken from $\rho$ the AtomWork \cite{Xu-JJAP-2011} database.
\end{tablenotes}
\label{table:MCEs}
\end{threeparttable}
\normalsize

\newpage

\newpage
\setcounter{figure}{0}
\renewcommand{\figurename}{Figure S.}

\begin{figure}[t]
\centering
\includegraphics[width=120mm]{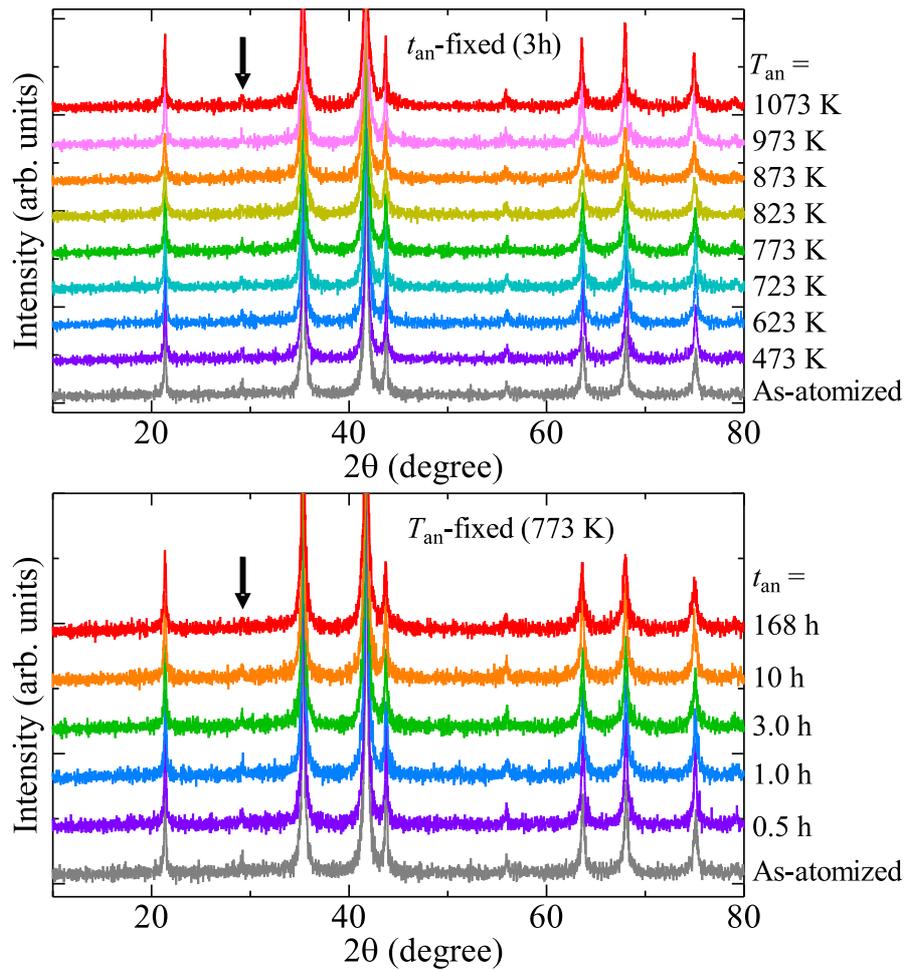}
\caption{(Color Online)
Powder XRD patterns of the ErCo$_{2}$ atomized particles
for $t_{\rm an}$-fixed (the upper panel) and $T_{\rm an}$-fixed (the lower panel).
The arrows represent the Bragg peak position of Er$_{2}$O$_{3}$.} 
\label{fig:S1}
\end{figure}

\begin{figure}[t]
\centering
\includegraphics[width=130mm]{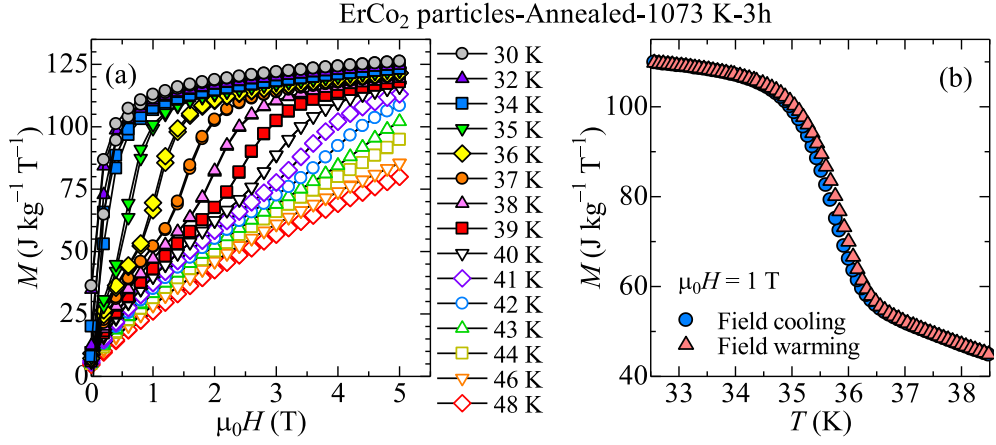}
\caption{(Color Online)
Magnetic properties of ErCo$_{2}$ atomized particles annealed at 1073 K for 3h.
(a) $M$-$H$ curves at various temperature near $T_{\rm C}$.
(b) $M$-$T$ curves at 1 T under field cooling and field warming processes.} 
\label{fig:S2}
\end{figure}

\begin{figure}[t]
\centering
\includegraphics[width=130mm]{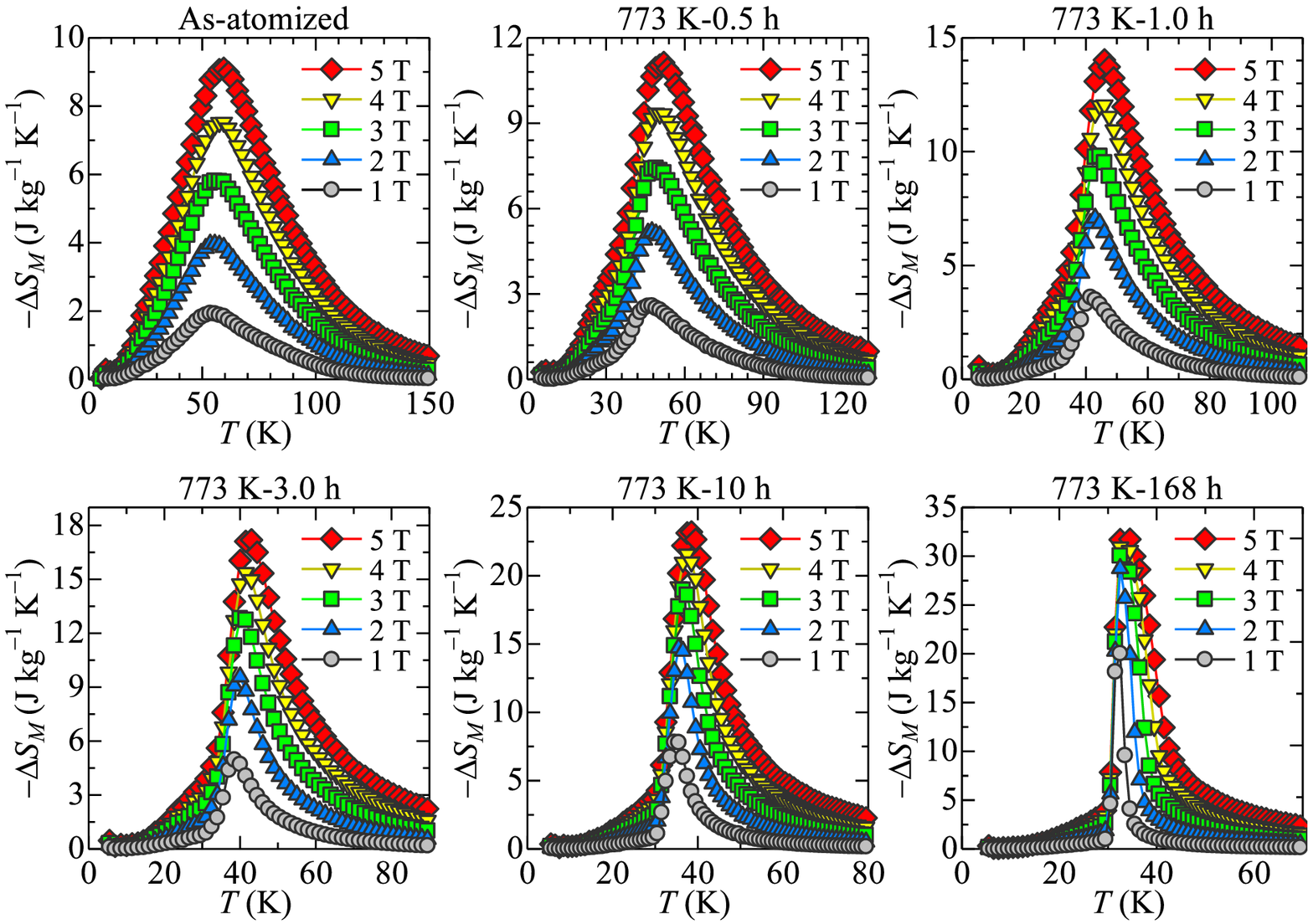}
\caption{(Color Online)
Temperature dependence of $-\Delta S_{M}$ of the ErCo$_{2}$ atomized particles
in case of $T_{\rm an}$-fixed for various $\mu_{0}\Delta H$.} 
\label{fig:S3}
\end{figure}

\begin{figure}[t]
\centering
\includegraphics[width=130mm]{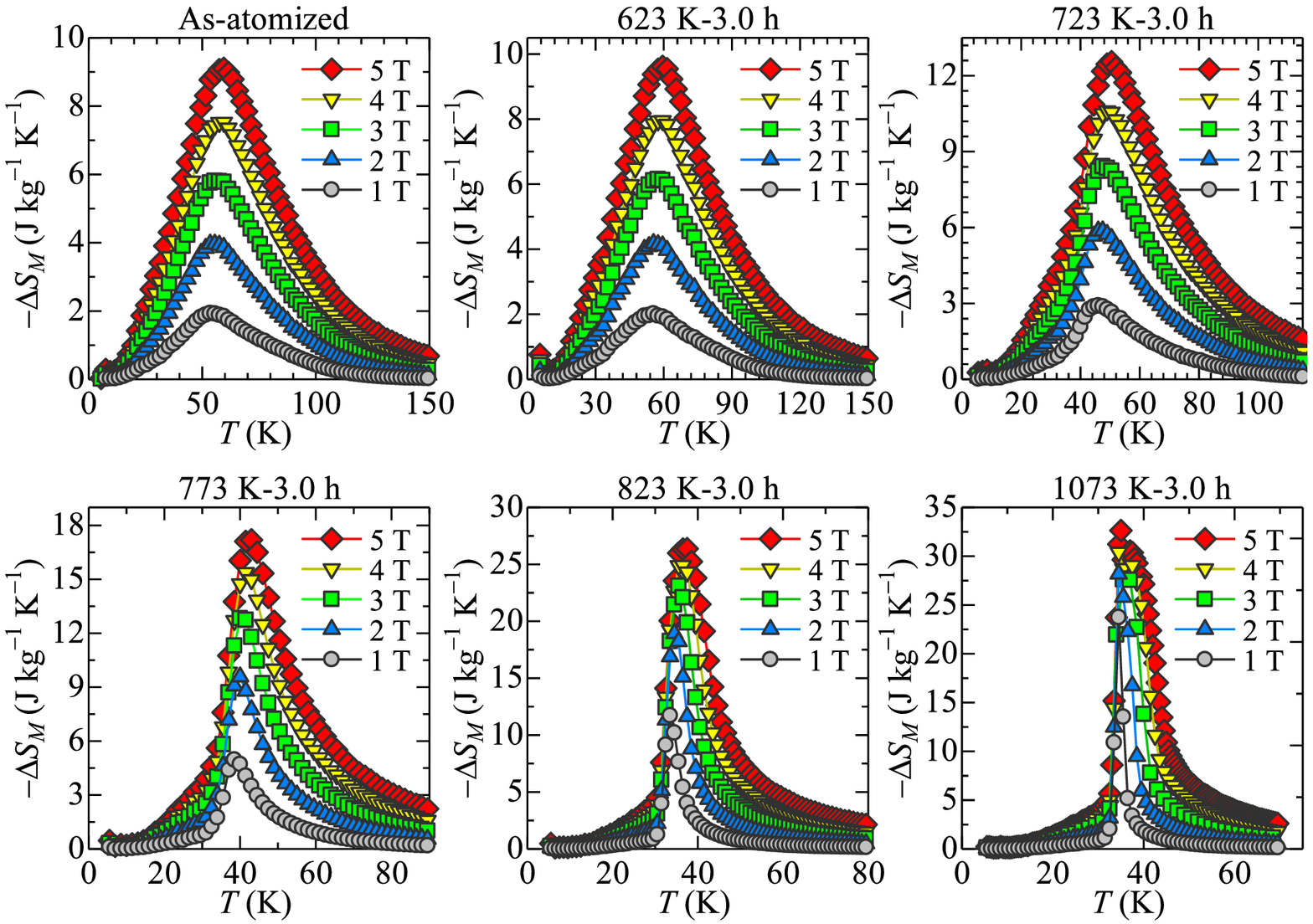}
\caption{(Color Online)
Temperature dependence of $-\Delta S_{M}$ of the ErCo$_{2}$ atomized particles
in case of $t_{\rm an}$-fixed for various $\mu_{0}\Delta H$.} 
\label{fig:S4}
\end{figure}

\end{document}